\documentstyle[12pt,amssymbols]{article}

\newcommand{\CP}{{\Bbb C}{\bf P}}
\begin{document}

\begin{titlepage}
\null \vskip -0.6cm
\today
\null \hfill \parbox{4cm}{PAR--LPTHE 92--17

}

\vskip 1.4truecm
\begin{center}
\large DEFORMATIONS OF DYNAMICS

ASSOCIATED TO THE CHIRAL POTTS MODEL
\end{center}
\vskip 1truecm
\centerline{M.P. Bellon\footnote{
Present address: Laboratoire de Physique Th\'eorique ENSLAPP, Ecole Normale
Sup\'erieure de Lyon, 46 All\'ee d'Italie, F-69364 Lyon Cedex 07, FRANCE
}, J-M. Maillard, G. Rollet, C-M. Viallet}
\vfill
\vfill

\noindent {\bf Abstract.}

We describe deformations of non-linear (birational) representations of
discrete groups generated by involutions, having their origin in the
theory of the symmetric five-state Potts model.
One of the deformation parameters can be seen as the number $q$ of
states of a chiral Potts models.  This analogy becomes exact when $q$ is a
Fermat number.  We analyze the stability of the corresponding dynamics, with
a particular attention to orbits of finite order.

\vfill
 {\raggedright
\noindent PACS: 05.50, 05.20, 02.10, 02.20 \par
\noindent AMS classification scheme numbers:
        82A68, 82A69, 14E05, 14J50, 16A46, 16A24, 11D41 \par
\medskip
\noindent {\bf Key-words}: Birational transformations,
star-triangle equations, Inversion relations, Integrable models, Iterated
mappings, Automorphisms of algebraic curves, \nobreak{Dynamical} systems,
Elliptic curves, Cremona transformations, Fermat numbers, cyclotomic
polynomial.\par }\vskip 1cm
\centerline{work supported by CNRS}
\vskip .5truecm
\hrule\vskip .5truecm
\begin{center}
\obeylines
Postal address: Laboratoire de Physique Th\'eorique et des Hautes
Energies Universit\'e de Paris VI et VII, URA280 Tour 16, $1^{\rm er}$ \'etage,
bo\^\i te
126.  4 Place Jussieu/ F--75252 PARIS Cedex 05
\end{center}
\end{titlepage}

\section{Introduction}
\null\indent
In previous publications~\cite{bmv1,bmv1b,bmv2,bmv2b,prl}, we have presented a
class of {\it birational transformations} in projective space, which are
symmetries of the Yang-Baxter equations (or star-triangle relation and their
higher dimensional generalizations) and also
symmetries for the model.  These  transformations are
generated by involutions which represent the {\it inversion relations}
of lattice models of statistical mechanics~\cite{Ba81,Ba82,St79}.  They
amount to taking inverse of matrices whose entries belong to the
parameter space.

This construction has a priori a certain rigidity: it provides with
explicit examples of birational automorphisms of $\CP_N$
for any $N$, with {\it integer} coefficients. Some of these mappings
have non-trivial properties : their orbits are dense in non trivial algebraic
subvarieties of $\CP_N$.

We study here some deformations of the  transformations appearing
in the  symmetric five-state Potts model, for which the parameter space
  identifies with projective space $\CP_2$.
We introduce families of deformations, depending on respectively   four, two
and one parameters.
For the  single--parameter family, this
parameter $q$ can be seen heuristically as the number of colours of
$q$-state chiral Potts models,  and the orbits under iteration of the
transformations {\it lie on algebraic curves}
(this is no longer the case for the two (or four) parameter families of
deformations we produce).
 We thus have a {\it $q$-deformation} of the linear pencil of elliptic curves
foliating the parameter space $\CP_2$ of the symmetric five state Potts model.
  Among the orbits, the ones where
the group has a {\it finite order} representation pop out immediately, and we
describe some of them.

\section{Construction} \label{construction}
\null
\indent
Let us consider $q$-state spin models with nearest neighbour interactions on
a square lattice. The Boltzmann weights $w(\sigma_i,\sigma_j)$ for an
oriented bond $\langle i j \rangle$ can be seen as the entries $m_{ij}$ of
a $q \times q$ matrix.
 The matrix of Boltzmann weights is a $q\times q$ matrix $M$
with (complex) homogeneous entries $m_{ij}$. Choosing a specific model means
fixing $q$ and imposing constraints on the Boltzmann weights.

\subsection{The inversion relation}

Two distinct inverses act on the matrix of Boltzmann weights: the matrix
inverse $I$ and the Hadamard (element by element) inverse $J$. Let us write
down the inversion relations~\cite {Ba80,Ba82} :
\begin{eqnarray}
\label{i1}
         \sum_\sigma w(\sigma_i,\sigma) \cdot I(w)(\sigma,\sigma_j) &=& \mu
\; \delta_{\sigma_i \sigma_j}, \\ w(\sigma_i,\sigma_j) \cdot
J(w)(\sigma_i,\sigma_j) &=& 1. \label{i2}
\end{eqnarray}
where $ \delta_{\sigma_i \sigma_j}$ denotes the usual Kronecker delta.
This gives the following action on the matrix $M$:
\begin{eqnarray}
        I&:&\quad M \longrightarrow M^{-1} \\ J&:&\quad m_{ij}
\longrightarrow 1/ m_{ij}
\end{eqnarray}
In the inversion relations of two dimensional lattice models of
Statistical Mechanics~\cite{Ba82,St79,Ba80,Ma83,HaMaRu89}, one acts on the
Boltzmann weights for vertical bonds with $I$ (resp.~$J$) and on the
horizontal bonds with $J$ (resp.~$I$).

The two involutions $I$ and $J$ generate an infinite discrete group
$\Gamma$ isomorphic to the infinite dihedral group ${\Bbb Z}_2 \ltimes
{\Bbb Z}$.  The ${\Bbb Z}$ part of $\Gamma$ is generated by $IJ$.  In
the parameter space of the model, that is to say some projective space
$\CP_{N-1}$ ($N$ homogeneous parameters), $I$ and $J$ are (bi)rational
involutions.  They give a {\em non-linear representation} of this group
by an infinite set of birational transformations~\cite{bmv1}.  It may
happen that the action of $\Gamma$ on specific points yields finite
orbits (the representation of $\Gamma$ identifies with the $p$-dihedral
group ${\Bbb Z}_2 \ltimes {\Bbb Z}_p$). One only wants to consider
constraints between the Boltzmann weights preserved by these two
inversions.

We choose to  restrict ourselves to constraints on the matrix of Boltzmann
weights, of the form $m_{ij}=m_{kl}$ for a number of pairs of indices.
Such constraints are automatically preserved by $J$. They amount to
giving a partition of the set of the entries of $M$ such that all
elements of a given part are set equal.  Only a limited number of
partitions give a pattern such that the matrix inverse $I$ will
preserve it.  We call these patterns {\it admissible}
patterns~\cite{bmv4}.

On these remaining parameters $x_0,x_1,\dots,x_{N-1}$, (that is to say a
point in $\CP_{N-1}$), the action of the Hadamard inverse $J$ is simply $ x_k
\longrightarrow 1/x_k$.  The action of the matrix inverse $I$ is $x_k
\longrightarrow i_k(x_0,x_1,\dots,x_{N-1})$, where the $i_k$ are
polynomials with integer coefficients in the $x_j$'s.

The matrix of Boltzmann weights for the $q$-state chiral Potts model is
the  general cyclic $q\times q$ matrix:
\begin{equation}
\label{x_0 x_q}
M =  \pmatrix { x_0	& x_1	& x_2	&	&\ldots&x_{q-1} \cr
                x_{q-1} & x_0	& x_1	& x_2	&\ldots&x_{q-2}\cr
                x_{q-2} & x_{q-1}	& x_0	& x_1 \cr
				& x_{q-2} & x_{q-1} & x_0 & \ddots &\vdots\cr
		    \vdots  & \vdots	&	& \ddots & \ddots & x_1\cr
			x_1	& x_2   & \ldots & \ldots & x_{q-1} & x_0 \cr
        }.
\end{equation}

For many models of statistical mechanics on lattices, the matrix of
Boltzmann weights is a stochastic matrix ($\sum_{j}m_{ij}=$independent
of $i$). We have remarked that this is the case of our
``admissible patterns'' for the classifications we have
performed~\cite{bmv1,bmv4}).
One remarks that our matrix algebra identifies with Bose-Mesner algebra
occurring in algebraic combinatorics~\cite{Ja91,BrCoNe89,BaIt84}.

\subsection{Orbits}
\label{orbits}
When the iteration of $IJ$ leads to algebraic curves, there exist
an algebraic invariant $\Delta$ such that the curve's equation  reads
$\Delta=\mbox{\it cste} $~\cite{bmv1}.
  As seen in~\cite{bmv6}, there is  an elliptic uniformization~\cite{Ma86},
making $IJ$ a mere translation $\theta \rightarrow \theta +\lambda$ of
the uniformizing (spectral) parameter  $\theta$.
The situation is that one has translations on the circle $S^1$ with a shift
not commensurate to the circumference (irrational `rotation number').
The generic orbits are thus dense in the (algebraic) curves.

\subsection{Symmetric five-state Potts model}

The symmetric five-state Potts model~\cite {bmv2b} corresponds to matrix
(\ref{x_0 x_q}) for $q=5$ when it is a symmetric one.
{}From the very construction of the two involutions $I$ and $J$, they are
birational transformations with {\em integer} coefficients.  It is  possible
to introduce deformations of our mappings. We will  concentrate here on
deformations of the mappings associated with the symmetric five-state
chiral Potts model. The  transformations $I$ and $J$ read~\cite{bmv2},
in terms of  the inhomogeneous variables $u=x_1/x_0$,  and  $v=x_2/x_0$:
\begin{eqnarray}
        I:\quad u &\rightarrow& { -u+uv-u^2+v^2 \over 1 +u +v-v u -v^2 -u^2 }
	   \qquad v \rightarrow { -v+uv-v^2+u^2 \over 1 +u +v-v u -v^2 -u^2 } ,
\label{IZ5} \\ &&\nonumber \\
	J:\quad u &\rightarrow& 1/u, \qquad v\rightarrow 1/v. \label{JZ5}
\end{eqnarray}

In the examples developed
in~\cite{bmv1}, there are collineations intertwining the two involutions
(matrix inverse $I$ and element by element inverse $J$).  They
generalize the Kramers-Wannier duality~\cite{bmv1,bmv1b}.  This
property is not too surprising since in $\CP_2$, the Noether
theorem~\cite{Sh77} proves that
every birational automorphism of the plane can be represented as a product
of quadratic transformations and projective transformations.
The birational transformations in $\CP_n$, $n>2$, are much more complicated.

The collineation intertwining $I$ et $J$ for this model is an involution
and reads:
\begin{eqnarray}
\label{KW}
D:\quad \quad u &\rightarrow&{1+(\omega+\omega^4)u+(\omega^2+\omega^3)v \over
1+2\,u+2\,v}\quad,
\nonumber \\
v &\rightarrow&{1+(\omega^2+\omega^3)u+(\omega+\omega^4)v \over 1+2\,u+2\,v}
\end{eqnarray}
with $\omega=\exp ( 2 i \pi / 5)$.

\subsection{Invariants}
\label{invariants}
\null\indent
The localization of the orbits of $\Gamma$ on curves, or more generally
non trivial subvarieties of the space of parameters, rather than as
``clouds'' of points, is the sign of the existence of {\it algebraic}
invariants of the group $\Gamma$~\cite{bmv1}.  Equating these
invariants to some constant gives the equations of the algebraic
varieties.  For the symmetric five-state Potts model a linear pencil of
algebraic curves (elliptic curves) emerges in the study of the orbits
of the group $\Gamma$ generated by $I$ and $J$~\cite{bmv1}.  The
rational expression
\begin{eqnarray} \label{delta0}
	\Delta = {(u^2 + v^2 +3uv)(u-1)(v-1) \over
		(u-v)^{2}(2+3\,u+3\,v+2\,uv) },
\end{eqnarray}
is invariant by $I$ and $J$.  The curves in the pencil are given by
$\Delta=const$. This can be seen on the orbits of $\Gamma$~\cite{bmv1},
and verified explicitly by a direct calculation. One has a linear
pencil of curves, with intersection points at which the invariant
$\Delta$ is undetermined. This means that our mappings are defined in
$\CP_2$ {\it minus} some points.

One sees that $\Gamma$ is an {\it infinite set of automorphisms} of the
subvarieties.  Therefore, if the subvarieties are curves, they are
necessarily of genus 0 or 1~\cite{Ma86}.  A detailed analysis to be
found in parallel publications~\cite{bmv6} shows that one has
pencils of curves which are generically of genus 1.  For a {\it finite
number} of values of $\Delta$, the curve factorizes in a number of
genus 0 components with a rational parametrization.  It is clearly the
case for the line $u=v$, associated with the standard scalar Potts
limit, which is stable by $\Gamma$ and for which one can introduce a
rational parametrization to represent $\Gamma$~\cite{JaMa82}. Actually,
in this model, this line $u=v$ comes together with an hyperbola of
equation:
\begin{eqnarray}
\label{hyper}
        2uv + 3 (u+v) +2 = 0.
\end{eqnarray}
Numerically, the line $u=v$ is not stable and the iterations of $IJ$
escape to the hyperbola at the points $u=v={1\over 2}(-3\pm \sqrt{5})$
\cite{bmv1b}. One has genus zero curves {\it only} for the following
values of $\Delta$: $\infty$, $-1/8$, 0, and ${1\over 2}(11\pm 5\sqrt{5})$.

Notice that the points of the curve $\Delta=1/2$ have a finite orbit of
order $6$ under $\Gamma$.  More generally, subvarieties made out of points
having finite orbits are automatically stable.  For example the integrability
variety of Au-Yang et al. is such a variety~\cite{HaMa88,HaMaRu89}.
More examples are easy to find explicitly by writing the condition
$(IJ)^r=identity$, for arbitrary integer $r$. One guesses that all these
finite orbit subvarieties belong to the linear pencil. The curves
$\Delta={1\over 2}(11\pm 5\sqrt{5})$ decompose into the hyperbolae~\cite
{bmv2b}:
\begin{eqnarray}
\label{hyperor}
        (u^2 -v )(\omega^{ \pm 2} + \omega^{ \pm 3}) -u(1-v) =0,\\
        (v^2 -u )(\omega^{ \pm 2} + \omega^{ \pm 3}) -v(1-u) =0,
\end{eqnarray}
with $\omega=\exp ( 2 i \pi / 5)$.  Acting on the
points of these curves, $IJ$ is of order five and the action of
$\Gamma$ gives finite orbits.  Also notice that $\Delta$ is invariant
by the collineation (\ref{KW}) which is nothing but the Kramers-Wannier
duality transformations, that is to say the Fourier transform in ${\Bbb
Z}_5$~\cite{Bi76,Bi77}).

\subsection {Deformations of the birational transformations}

We may look for more general collineations by not demanding that they are
involutive.  We request that they send $(0,0)$ onto $(1,1)$ since
this constraint has a natural physical meaning: it is the existence of
a mapping from the
(high-temperature) decoupling limit where the matrix (\ref{x_0 x_q})
tends to the constant matrix, to the (low-temperature) limit where
this matrix tends to the identity matrix. One thus gets a four
parameter family of collineations which are generically of infinite
order.

They read
\begin{eqnarray}
\label{ABCD}
&u&\rightarrow{1-v+A(u-v) \over 1+Cu+Dv} \nonumber \\
\nonumber \\
and \quad &v&\rightarrow{1-u+B(u-v) \over 1+Cu+Dv}
\end{eqnarray}

The collineation associated to spin models must be such that one recovers
the duality transformation for the standard scalar Potts model $u=v$. This
gives a further constraint on C and D :
\begin{eqnarray}
\label{CD}
C+D=q-1
\end{eqnarray}

The previous birational transformations can be deformed  by breaking the
symmetry between the two (inhomogeneous) variables u and v. The
collineations which are the duality transformations
of some $q$-state nearest neighbour spin model~\cite{Bi77}, satisfy
the following conditions : they are { \it involutions or
transformations of order four}~\cite {HaMa88,HaMaRu89} and they map
the point $(u,v)=(0,0)$ onto the point $(1,1)$.  This last constraint
corresponds to the low-to-high temperature correspondence (it maps the
identity $q \times q$ matrix onto the matrix were all the entries are
equal to one which corresponds to the decoupling of the spins). These
constraints are quite severe. There is actually a two-parameter family
of such collineations $D_{Q,R}$:
\begin{eqnarray}
\label{DQR}
	D_{Q,R}:\quad u&\to& \widehat u
		={1-(u+v)/2+(Q+R)(u-v)/2\over 1+(Q^2-1)(u+v)/2-R(Q-1)(u-v)/2}
\nonumber \\ \nonumber \\
	v&\to &\widehat v
		={1-(u+v)/2-(Q-R)(u-v)/2\over 1+(Q^2-1)(u+v)/2-R(Q-1)(u-v)/2}
\end{eqnarray}

It is straightforward to look at the orbits of $I_{Q,R}J=D_{Q,R}JD_{Q,R}J$ and
see that these orbits generically {\it do not preserve }algebraic curves.

However various lines are preserved: the line $u=v$ and the set of
lines $u=1$, $v=1$, $\widehat u =1$, $\widehat v =1$.

\section{$q$-generalizations}
The duality transformation for such general $q$-state chiral Potts model read
on the $q$ homogeneous parameters of the model~\cite{HaMa88,HaMaRu89}
\begin{eqnarray}
\label{dual}
 x_i\to\widehat x_j=\sum_{\alpha =0}^{q-1}\omega^{\alpha j}.x_\alpha
\end{eqnarray}
Let us restrict the model by imposing equalities between
the $x_i$'s, $i\not= 0$, such that these equalities are invariant under
the duality transformation (\ref{dual}) and thus under the group $\Gamma$
(they are automatically invariant by the Hadamard inverse J).

In terms of the remaining inhomogeneous variables $y_i=x_i/x_0$,
$i=i_1,\ldots,i_{N-1}$, transformation (\ref{dual}) is a collineation.
Let us concentrate on the cases where there are only two inhomogeneous
variables denoted $u$ and $v$. It then reads :
\begin{eqnarray}
\label{dual2}
	D:\quad \quad u&\to&\widehat u={1+\alpha u+\beta v\over 1+(q-1)(u+v)/2}
\nonumber \\
		v&\to&\widehat v={1+\alpha v+\beta u\over 1+(q-1)(u+v)/2},
\end{eqnarray}
where $\alpha$ is a sum of power of $q^{th}$ root of unity, $\beta$
being the sum of the other power of $q^{th}$ root of unity, $\alpha$ and
$\beta$ satisfy $\alpha + \beta =-1$.

The ``duality'' transformation (\ref{dual}) is a transformation { \it
of order four}~\cite{HaMa88,HaMaRu89}. This implies the following
relation on $\alpha$ and $\beta$:
\begin{eqnarray}
\label{relation}
	-4\,\alpha\beta=\pm q-1 \nonumber
\end{eqnarray}
It is an involution if and only if the $q \times q$ matrix
(\ref{x_0 x_q}) is symmetric ($x_i=x_{q-i}$) which read on $\alpha$ and
$\beta$ : $-4\alpha\beta =q-1$ and gives the exact expression of $\alpha$
or $\beta$ :
\begin{eqnarray}
\label{alpha}
\alpha ={-1\pm \sqrt{q}\over 2},\quad \beta ={-1\mp \sqrt{q}\over 2}
\end{eqnarray}
Transformation (\ref{dual2}) with $\alpha$ and $\beta$ given by
(\ref{alpha}) correspond to  the $R\to 0$ limit of (\ref{DQR}) with $Q^2=q$.

It is an interesting question to see for which value of the integer $q$
(different from $q=5$) the values ${1\over 2}(-1\pm \sqrt{q})$ can be
realized as sum of $q^{th}$ root of unity. This question will not be
addressed here.  However recalling Gauss, Kummer and Vandermonde's ideas on
cyclotomic polynomials one can see for instance that $q=17$ is such a
number.  We have the following identities involving $17^{th}$ root of unity:
\begin{eqnarray}
\label{cyclo1}
\omega^{3^0}+\omega^{3^2}+\omega^{3^4}+\omega^{3^6}+ \ldots+ \omega^{3^{14}}
	&=& {\textstyle {1\over2}}(-1+ \sqrt{17})/2 ,\\
\label{cyclo2}
\omega^{3^1}+\omega^{3^3}+\omega^{3^5}+\omega^{3^7}+ \ldots+ \omega^{3^{15}}
	&=& {\textstyle {1\over2}}(-1- \sqrt{17})/2 .
\end{eqnarray}
These special values of $q$ correspond to {\it Fermat numbers}\footnote{Fermat
	numbers are closely connected with problem of geometric
	construction of regular polygons by means of ruler and compass.
	For the construction of a regular $n$-gon to be possible it is
	necessary, and sufficient, that the representation of n as a
	product of prime numbers takes the form $n=2^N\,p_1 \ldots p_k$
	where $N \geq 0,\,p_1,\ldots, p_k$ are all different prime
	numbers of the form $2^M+1$. This geometric construction is in
	one-to-one correspondence with the resolution of the equation
	$x^q-1=0$ (the $q$-gon) in terms of a ``tower'' of roots of
	quadratic equations corresponding to the subsums of $q^{th}$
	root of unity similar to (\ref{cyclo1}) and (\ref{cyclo2})}
(integers of the form $2^{2^{r}}+1$) or products of such numbers up to
some power of 2.  This underlines some particular values of $q$ for the
$q$-state nearest neighbour Potts models: $q=3$, 5, 17, 257, 65537\dots

Let us consider the group $\Gamma$ generated by the two involutions $I=DJD$
and $J$, where $D$ is given by (\ref{dual2}).  The orbit of
$\Gamma$ lie remarkably on an algebraic curve (see figure~1).

These algebraic curves are given by the equation $\Delta=$constant, where
$\Delta$ reads :
\begin{eqnarray}
\label{deltaq}
	\Delta={(u-1)(v-1)((q-1)(u^2+v^2)+2(q+1)uv)\over(2+(q-2)(u+v)+2uv)(u-v)^2}
\end{eqnarray}
These algebraic curves form a {\it linear pencil of elliptic curves},
which generalize the linear pencil of elliptic curves already obtained for
the symmetric five-state Potts model~\cite{bmv1}(and equation
(\ref{delta0})). One notes that $\Delta$ is invariant by $D$ given by
(\ref{dual2}) and (\ref{alpha}). These curves are generically elliptic
curves except for a {\it finite number} of values of $\Delta$. $\Delta=0$ is
clearly made out of genus zero curves. The vanishing of $\Delta$ corresponds to
four lines $u=1$, $v=1$, $u=c_q v$ (which is
nothing but $\widehat u-1=0$) and $v=c_q u$, where:
\begin{eqnarray}
\label{C_q}
	c_q={-(q+1) + 2\, \sqrt{q} \over q-1}.
\end{eqnarray}
The line $u=v$, which corresponds to the standard scalar Potts model, is
known to have a rational parametrization~\cite{JaMa82}. The hyperbola
appearing  in the denominator of $\Delta$ is also a genus zero curve.
Note that it is not only $\Gamma$ invariant but that is also (globally)
invariant by the duality transformation (\ref{dual2}). $\Delta=-1/2$
also correspond a curve of  genus zero  (two lines and one hyperbola) {\it
remarkably independent of the value of $q$} :
\begin{eqnarray}
\label{noq}
(u+v)(u+v-2)(2uv-u-v)=0
\end{eqnarray}

The exhaustive list of values of $\Delta$ for which the curve is of genus zero
is completed with the values:
\begin{equation}
 	\Delta = -2 { 1 \mp \sqrt{q} \over (2 \mp \sqrt{q})^2 }
\end{equation}
the equation of the curve being
\begin{eqnarray}
\label{del}
(u+v-v^2-uv)\sqrt{q}\pm (v^2-u-3v+3uv)=0
\end{eqnarray}
and of course the equation obtained exchanging $u$ and $v$.

Eq.~(\ref{del}) generalize the two equations already obtained for the symmetric
five-state Potts model~\cite{bmv1} (equation (\ref{hyperor})). These
equations were finite orbits of $IJ$ for $q=5$ ($(IJ)^5=e$). This is no
longer the case for arbitrary $q$ :  the orbits of the points of
equation (\ref{del}) are of infinite order. On the other hand, finite
orbits for the group $\Gamma$ generated by transformations
(\ref{dual2}) can easily be exhibited for arbitrary $q$. For instance
the following two polynomials are respectively the curves of the points
of order three and four:
\begin{eqnarray}
\label{order3}
8\,uv(v-1)(u-1)-(u+v)(u-v)^{2}(q-1) = 0
\end{eqnarray}
and
\begin{eqnarray}
\label{order4}
	(u^3vq - 4\,qu^2v+2\,v^2qu^2+qu^2+2\,quv - 4\,quv^2+uv^3q+qv^2 - v^2
\nonumber \\
	-6\,uv+8\,uv^2 - uv^3 - u^2+8\,u^2v - 6\,u^2v^2 - u^3v)
\nonumber \\
	(-4\,uv - 2\,u^2 - 2\,v^2+u^3 - 2\,u^3q+u^3q^2-4\,u^2v^2+7\,u^2v+7\,uv^2
\nonumber \\
	+2\,quv^2+2\,qu^2v-4\,quv+2\,qu^2+2\,qv^2-2\,uv^3-2\,u^3v+2\,u^3vq
\nonumber \\
	+2\,uv^3q-4\,v^2qu^2+v^3-2\,v^3q+v^3q^2-uq^2v^2-vq^2u^2)&=& 0.
\end{eqnarray}
The curve of the points of order 3 is the curve $\Delta = {1\over 2}(q-1)$, and
the points of order 4 are on the product of the curves $\Delta = - (q-1)/(q-2)$
and $\Delta = {1\over 2} (q^2 -1)$.
The equation for the points of order 5 is the product of two polynomials of
degree 6 in $u$ and $v$, of degree 3 in $q$ and each of them is the sum of
94 monomials.  They can be seen to be the product of the equations of two
curves of the pencil, with conjugated values of $\Delta$ :
\begin{eqnarray}
	\label{order5}
	\Delta &=& -{1\over 4} (q-1) ( q^2 - q + 2 \pm q\sqrt{q^2 -2q +5}),\\
	\Delta &=& -{1\over 2} { ( 1 \pm \sqrt{q} ) ( 1 \mp \sqrt{q})^2) \over q
		-1 \pm \sqrt{q} }.
\end{eqnarray}
For $q=5$ the polynomial corresponding to the values of $\Delta$
(\ref{order5}) factorizes and one recovers equations (\ref{del}) :
\begin{eqnarray}
\label{P}
P(u,v,q)=&(u^4-u^3-u^2+u^3v+uv-u^2v^2-uv^2+v^2) \nonumber \\
&\quad (v^4-v^3-v^2+v^3u+uv-u^2v^2-u^2v+u^2)
\end{eqnarray}
The study of the points of order 6 yields also a product of curves from the
pencil, with the 4 values of $\Delta$:
\begin{equation}
	\Delta = - {1\over2} {(q -1 )(q+3) \over q-3 },
\end{equation}
and the three solutions of the following order 3 polynomial:
\begin{equation}
	x^3 - {1\over2}(q-1)(q^3-3q^2+3q+3)x^2 - {3\over4}(q^2-2q-1)(q-1)^2x
		- {1\over8}(3q+1)(q-1)^3.
\end{equation}

\noindent
{\bf Remark:} the other one-parameter dependent collineation (\ref{dual2}) of
order {\it four} corresponds to:
\begin{eqnarray}
\label{alphai}
	\alpha=(-1 \pm i \sqrt{q})/2 \quad \mbox{and} \quad
		\beta=(-1 \mp i \sqrt{q})/2.
\end{eqnarray}
However the group $\Gamma$ generated by the two involutions $J$ and
$I=DJD^{-1}$ where $D$ is given by (\ref{dual2}) for $\alpha$ and $\beta$
given by (\ref{alphai}), do not lead to a linear pencil of curves.
Again one can ask the question to see for which value of the integer
$q$ the values (\ref{alphai}) can be realized as sum of $q^{th}$ root
of unity. Actually $q=7$ is such an integer with $\alpha=\omega +\omega
^{2}+\omega ^{4}$, $\beta=\omega ^{3} +\omega ^{5}+\omega ^{6}$, where
$\omega ^{7}=1$.  This model has been introduced elsewhere (it is
referred to model ${\Bbb P}_{4}$ in~\cite{bmv1,bmv1b}). The orbits of
$\Gamma$ for this model are quite chaotic (see figure~4 in~\cite{bmv1}
and figure~6 in~\cite{bmv1b}).

\section{Generalizations and outlook.}
All this analysis on birational transformations on $\CP_2$ can be
generalized straightforwardly to $\CP_n$ ($n\geq 3$).  Our
transformations are generated by involutions in $\CP_n$.   In the most
general framework an exhaustive classification of the involutions in
$\CP_2$ has been performed by Bertini~\cite{Be1877,Di74}.

For involutions in $\CP_n$ the situation is much more
involved: there is no longer Noether-Castelnuovo theorem imposing each
Cremona transformation to be a finite product of quadratic
transformations~\cite{GrHa78}.

It will be necessary to restrict the analysis to particular classes of
birational transformations.  A guide could be their links with
statistical mechanics on lattices and more precisely integrable models of
nearest neighbour interaction spin models.   We think in particular
 that an interesting example is the group of transformations in $\CP_3$
generated by the Hadamard inverse $J$ and $I=DJD$, where $D$ is an involutive
collineation mapping the point $(0,0,0)$ in $(1,1,1)$.

\bigskip
{\bf Acknowledgments}: We would like to thank J.~Avan and S.~Boukraa
for very stimulating discussions and comments.


\end{document}